\documentclass[aps,nofootinbib,showpacs,twocolumn]{revtex4-1}
\usepackage[CJKbookmarks, pdftex, bookmarksnumbered, bookmarksopen, colorlinks, citecolor=blue, linkcolor=blue]{hyperref}

\usepackage[english]{babel}
\usepackage{amstext,amsmath,amssymb,amsfonts,bbm}
\usepackage[latin1]{inputenc}
\usepackage{graphicx}
\usepackage{epstopdf}
\usepackage{amsthm}
\usepackage{tocvsec2}
\usepackage{enumerate}
\usepackage{subfigure}

\usepackage{bm}
\usepackage{color}
\usepackage{stmaryrd}

\usepackage{amsthm}
\usepackage{amsmath}
\usepackage{mathtools}
\usepackage{tabularx}
\usepackage{accents}
\usepackage{bbold}
\usepackage{dsfont}
\usepackage{bm}

\newcommand{\Ref}[1]{(\ref{#1})}

\newcommand{\U}{\mathrm{U}}

\newcommand{\be}{\begin{equation}}
\newcommand{\ee}{\end{equation}}
\newcommand{\bea}{\begin{eqnarray}}
\newcommand{\eea}{\end{eqnarray}}

\newcommand{\bit}{\begin{itemize}}
	\newcommand{\eit}{\end{itemize}}

\newcommand{\tr}{{\rm Tr}}

\def\p{\partial}

\newcommand{\ex}{\mathrm{e}}
\newcommand{\dd}{\mathrm{d}}

\newcommand{\bra}[1]{\left\langle {#1}\right|}
\newcommand{\ket}[1]{\left|{#1}\right\rangle}
\newcommand{\braket}[2]{\langle {#1}|{#2}\rangle}

\let\m=\mu

\begin{document}

\title{Ryu-Takayanagi Formula for Symmetric Random Tensor Networks}

\author{Goffredo Chirco$^{1}$, Daniele Oriti$^{1}$, Mingyi Zhang$^{1}$}

\affiliation{$^1$Max Planck Institute for Gravitational Physics, Albert Einstein Institute, Am M\"{u}hlenberg 1, 14476, Potsdam, Germany.}

\date{\today}

\begin{abstract}
We consider the special case of Random Tensor Networks (RTN) endowed with gauge symmetry constraints on each tensor. We compute the R\`enyi entropy for such states and recover the Ryu-Takayanagi (RT) formula in the large bond regime. The result provides first of all an interesting new extension of the existing derivations of the RT formula for RTNs. Moreover, this extension of the RTN formalism brings it in direct relation with (tensorial) group field theories (and spin networks), and thus  provides new tools for realizing the tensor network/geometry duality in the context of background independent quantum gravity, and for importing quantum gravity tools in tensor network research.
\end{abstract}

\maketitle

\section{Introduction}
Tensor networks \cite{TN-intro} have developed into a powerful and ubiquitous formalism in quantum information and in the analysis of quantum many-body systems. It is first of all a very efficient way to capture the entanglement properties of such many-body systems, as well as quantum field theories (including lattice gauge theories), and it provides a general framework for describing (and identifying) quantum states characterized by area-laws, which indeed include the ground states of several interesting quantum many-body systems \cite{TN-results}. Somewhat surprisingly, it also provides a natural framework for investigating the nature of spacetime at the Planck scale. 
This comes about from two main directions.
First, different theoretical frameworks, from background independent quantum gravity to string theory, hint at a scenario, where continuum spacetime geometry is replaced, at a more fundamental level, by pre-geometric quantum degrees of freedom, often of purely combinatorial and algebraic nature. 
In (Tensorial) Group Field Theories (GFT) \cite{GFT} and random tensor models \cite{TM}, as well as in Loop Quantum Gravity (LQG) and Spin Foam Models \cite{LQG}, pre-geometric quantum degrees of freedom are encoded in \emph{random combinatorial network structures}, labelled by algebraic data. In particular, they are encoded in spin networks, graphs labeled by irreps of $SU(2)$ and endowed with a gauge symmetry at each node. These type of quantum states, in fact, are very close to tensor networks \cite{us}, and tensor network techniques have found already a number of quantum gravity applications \cite{TN-QG}. A discrete spacetime and geometry is naturally associated, at a semi-classical level, to such structures and their quantum dynamics is directly related to (non-commutative) discrete gravity path integrals \cite{SF-path}. The outstanding issue is then to show the emergence of {\it continuum} spacetime and geometry from the full quantum dynamics of the same pre-geometric degrees of freedom, which in fact describe a quantum spacetime as a peculiar sort of quantum many-body system \cite{continuum}. It is natural to expect that the entanglement between the fundamental entities constituting spacetime is crucial for its emergence and for understanding the quantum nature of geometry at the Planck scale, and thus that tensor networks techniques will provide relevant tools in this context.
Second, a different, but probably related relation between entanglement and geometry has been unraveled n the context of holographic gauge/gravity duality, and also here tensor network states recently acquired a prominent role. One key example is the Ryu-Takayanagi (RT) formula, which relates the entanglement entropy of a boundary region to the area of the minimal surface in the bulk homologous to the same region \cite{RT}.  The main intuition \cite{swingle} consists in interpreting the geometry of the auxiliary tensor network representation of a quantum many-body state as a representation of the dual spatial geometry. 

More recently, a lot of interest focussed on random tensor network (RTN) states, which are shown also to satisfy the Ryu-Takayanagi formula, as well as quantum error correction properties, in the large bond dimension limit \cite{hayden}. %For states with a semi-classical bulk dual, this allows to effectively map a generic many-body state of the boundary to a superposition of RTN's peaked around a classical geometry. 
The random character plays a central role in the study of the entanglement area laws in tensor networks. Indeed, random pure states are nearly maximally entangled states \cite{Page:1993df}, hence can be used as a toy model of a thermal state \cite{deutsch1991quantum, srednicki1994chaos}. This in particular implies that the computation of typical entropies and other quantities of interest for these states can be mapped to the evaluation of partition functions of classical statistical models \cite{hayden}. On the other hand, the interpretation of GFT fields as tensors provides an actual generalisation of the tensor network decomposition techniques, and any given GFT model would then provide tensor network states with a specific probability measure,  defined through its partition function Z and momenta (correlations), and tools for its evaluation

For the above reasons, we think it is very important not only to develop further all the above directions, but also to strengthen the links between these different uses of tensor networks.
The present work goes in this direction, by extending the formalism of RTN to incorporate one key feature of the random networks appearing in quantum gravity:  the \emph{gauge symmetry constraint} \cite{symmetricTN}, and deriving the RT formula in this extension.
%
% \\
%Spin networks, group field states in this sense, here we consider the case of a RTN with gauge symmetry a s a first step in this direction A d-dimensional GFT is a combinatorially non-local field theory living on (d copies of) a group manifold \cite{Oriti:2014uga,Baratin:2011aa,Oriti:2011jm,Oriti:2009wn}. Due to the defining combinatorial structure, the Feynman diagrams $\F$ of the theory are dual to cellular complexes, and the perturbative expansion of the quantum dynamics defines a sum over random lattices of (a prior) arbitrary topology. A similar lattice interpretation can be given to the quantum states of the theory. For GFT models where appropriate group theoretic data are used and specific properties are imposed on the states and quantum amplitudes, the same lattice structures can be understood in terms of simplicial geometries. The associated many-body description of such lattice states can be given in terms of a tensor network decomposition. The corresponding (generalized) tensor networks are thus provided with a field theoretic formulation and a quantum dynamics (and, in specific models, with additional symmetries). In this section, after a brief introduction to the GFT formalism, we detail this correspondence between GFT states and (generalized) tensor networks.

The paper is organized as follows. In the next section, we recall the definition of tensor network states, and of their random version. Then, we introduce the symmetric random tensor networks that we use in the rest of the paper. Having done so, we compute the $2$nd R\'{e}nyi entropy for random tensor networks endowed with a local gauge symmetry constraint and derive the RT formula for them. We end up with some concluding remarks.

\section{Tensor Network States and Holographic behaviour} 
A tensor network is a collection of tensors, associated to nodes of a network, connected by contractions operations, associated to links of the same network.

Generically, a  rank-$v$ (or $v$-valent) tensor ${T}$ is a multidimensional array of complex numbers with $v$ indices $\lambda$, each taking values from a set of dimension (\lq size\rq) $D_{|\lambda|} \in \mathbb{N}^+$ \cite{symmetricTN}. For simplicity, all indices are assumed to have the same size $D_{|\lambda|} = D$.

At the  quantum level, to each leg of the tensor one associates a Hilbert space $\mathbb{H}_D$, with dimension $D$, so that a  covariant tensor of rank $v$ is a multilinear form on the Hilbert space of the vertex ${T}\colon \mathbb{H}_n\equiv\mathbb{H}_D^{\otimes v}$.
Given an orthonormal basis $\ket{\lambda_n}$, $ n=1,\dots ,D$ in~$\mathbb{H}_D$, we can denote its components by
\begin{eqnarray}
\widehat{T}_{\lambda_1\cdots \lambda_d} \equiv { T} (\lambda_{1},\dots, \lambda_{v}),
\end{eqnarray}
hence generally define a \emph{tensor state} as
\begin{eqnarray}
|T\rangle = \sum_{\lambda_1,\dots \lambda_d} \widehat{T}_{\lambda_1\cdots \lambda_v}\ket{\lambda_1}\otimes\cdots\otimes\ket{\lambda_v} \qquad .
\end{eqnarray}
Graphically, we can represent any such tensor state as a vertex with $v$ open links emanating from it.
%A tensor network is simply given by a set of $d$-valent vertices $v$, corresponding to rank $d$ tensors, connected by contractions.

A state corresponding to a set of unconnected vertices is given by a tensor product of individual tensor states
\begin{eqnarray}
\ket{\mathcal{T}_{\mathcal{N}}}\equiv \bigotimes_n \ket{T_n} \quad.
\end{eqnarray}

In a connected network, individual tensor states are \emph{glued} by links, to each end of which we associate a Hilbert space $\mathbb{H}_D$. The Hilbert space of the link $\ell$ is then $\mathbb{H}_{\ell}=\mathbb{H}_D^{\otimes 2}$ while a link state can be written as
\begin{eqnarray}\label{tlink}
\ket{M}=M_{\lambda_1\lambda_2}\ket{\lambda_1}\otimes\ket{\lambda_2}
\end{eqnarray}
where the coefficients $M_{\lambda_1\lambda_2}$ indicate generic quantum correlations between the links ends.\footnote{One can observe it by defining a density matrix $\rho_M\equiv \ket{M}\bra{M}$ and tracing out one of the Hilbert space, without losing generality, tracing out $\mathbb{H}_D$ of $\ket{\lambda_2}$, then computing the von Neumann entropy of the reduced density matrix $\rho_1\equiv \text{Tr}_2\rho_M= M^{\dag}M$. The entropy $S= \text{Tr} \rho_1\ln \rho_1$ is non-zero unless $M_{\lambda_1\lambda_2}$ can split as $M_{\lambda_1\lambda_2}= A_{\lambda_1}B_{\lambda_2}$.  For simplicity, in the next sections we will often assume that the link state is maximally entangled, i.e. 
\begin{eqnarray}
\ket{M}=\frac{1}{\sqrt{D}}\delta_{\lambda_1 \lambda_2}\ket{\lambda_1}\otimes\ket{\lambda_2}.
\end{eqnarray}
The latter corresponds to the simplest type of gluing, imposing gauge invariance, in the spin networks states used in quantum gravity, and forming indeed a special type of tensor networks \cite{us, symmetricTN}.}
One could picture this gluing as the joining of two of the open links emanating from the original vertices (along their open ends), to form a link of the resulting network.

The entanglement of links encodes the information on the connectivity of the graph: two nodes are connected if their corresponding states contract with an entangled link state,
\begin{eqnarray}
\ket{\mathcal{T}_{12}}&\equiv& \bra{M}\ket{T_1}\ket{T_2}\\ \nonumber
 &=&
 T^{(1)}_{\lambda_1\cdots\lambda_a\cdots\lambda_v}\overline{M}_{\lambda_a\lambda_b}T^{(2)}_{\lambda'_1\cdots\lambda_b\cdots\lambda_u'} \bigotimes_{i\neq a}^v\ket{\lambda_i}\otimes\bigotimes_{j\neq b}^u\ket{\lambda'_i}
\end{eqnarray}
%Notice that if $\ket{M}$ was a non-entangled state, the connection would be trivial, i.e. the two nodes would be practically disconnected and the corresponding state could be written as a tensor product of two states,
%\begin{eqnarray}\nonumber
%\hat{\mathcal{T}}_{12}&=& T^{(1)}_{\lambda_a\lambda_1\cdots\lambda_v}\overline{A}_{\lambda_a}\bigotimes_{i=1}^v\ket{\lambda_i} \otimes \overline{B}_{\lambda_b}T^{(2)}_{\lambda_b\lambda'_1\cdots\lambda_u'}\bigotimes_{j=1}^u\ket{\lambda'_i}\\ 
%&=&\ket{T'_1}\otimes\ket{T_2'}
%\end{eqnarray}
Accordingly, given a network $\mathcal{N}$ with $N$ nodes and $L$ links, the corresponding \emph{tensor network state} is given by the contraction
\begin{equation}\label{eq:TNS}
\ket{\Psi_{\mathcal{N}}}\equiv \bigotimes_\ell^L\bra{M_{\ell}}\bigotimes_n^N\ket{T_n}\quad.
\end{equation}

As all but the open links of the network are contracted with nodes, $\ket{\Psi_{\mathcal{N}}}$ can be thought as an element of the Hilbert space $\mathbb{H}_{\partial\mathcal{N}}$ associated to the boundary (formed by the remaining open ends) of the network graph.

In lattice gauge theory, formulated in terms of tensor networks, the open links carry the physical degrees of freedom of the theory, while the fully contracted internal graph is interpreted as a virtual structure, whose correlation structure can be tuned to match the desired properties of the boundary lattice state $\ket{\Psi_{\mathcal{N}}}$ in $\mathbb{H}_{\partial\mathcal{N}}$. In quantum gravity formalisms based on spin networks (thus on tensor networks), like GFT and LQG, the internal graphs carry the degrees of freedom a spatial manifold, while the open links are associated to its boundary (corners of the spacetime manifold) and carry indeed additional degrees of freedom, related to the breaking of diffeomorphism symmetry on the same boundary \cite{open-links}. %Moreover, due to the intuition that the entropy of a tensor network is bounded by an area law agreeing with the Ryu-Takayanagi formula \cite{}  tensor network provide a natural framework for investigating the gauge theory gravity duality as well as for interpreting the geometry/entanglement correspondence via entanglement renormalisation algorithms \cite{}.

\subsection{Random Tensor Network States}

%Tensor networks can also be used to build holographic mappings or holographic codes [6Ð8], which are isometries between the Hilbert space of the bulk and that of the boundary.
Recently, a lot of interest has focussed on the study of networks of large-dimensional \emph{random} tensors (RTN). 

A convenient way to deal with RTNs is to consider the tensors $\widehat{T}_{\lambda_1\cdots\lambda_v}$ on each node are unit complex vectors $T_{\mu}$ chosen independently at random in their respective Hilbert spaces $\mathbb{H}_T \simeq \mathbb{H}_D^{\otimes v}$ (indeed, a $D^v$ dimensional vector space), with inner product $\overline{T}_{\mu}T'_{\mu}$. 
One can represent these vectors by choosing a gauge such that $T_\mu\equiv T_{\lambda_1\cdots \lambda_v}$, with 
\begin{eqnarray}\label{eq:gauge}
\mu = \sum_{a=1}^{v}\lambda_a D^{v-a}=0,1,\dots,D^4-1.
\end{eqnarray} 
Moreover, $\ket{T}\in \mathbb{H}_n$ being normalized, one has as well
\begin{eqnarray}
\braket{T}{T}=\overline{T}_{\lambda_1\cdots\lambda_v}T_{\lambda_1\cdots\lambda_v}=\overline{T}_\mu T_{\mu}=1
\end{eqnarray}

Notice also that the Hilbert space $\mathbb{H}_T$ corresponds to the fundamental representation of the group $\U(D^v)$.  
Given an arbitrary reference state $T^0_{\mu}$, a group element $U\in \U(D^v)$ will transform $T^0_{\mu}$ to a new vector $T^U_{\mu}\equiv (UT^0)_{\mu}$. A  random tensor $T_{\lambda_1\cdots\lambda_v}$ corresponds then to a random choice of the group element $U\in \U(D^v)$ defining $T^U_{\mu}$,\footnote{The random average of an arbitrary function $f (|Vx\rangle)$ of the state $|V_x\rangle$ is equivalent to an integration over $U$ according to the Haar probability measure $\int \dd U f(U\, |0_x\rangle)$, with normalization $\int \dd U = 1$. }  where the group element $U_n$ is independently chosen for each node of the network.\\

Idealised versions of RTNs, so-called pluperfect tensors, have been used to define bidirectional holographic codes, which simultaneously satisfies the Ryu-Takayanagi (RT) formula for a subset of boundary states, error correction properties of bulk local operators, a kind of bulk gauge invariance, and the possibility of sub-AdS locality.

More recently, in particular, building on the statistical properties of large dimensional random tensor network, the technique of \emph{random state averaging} was  used by \cite{hayden} to compute R\'enyi entropies and other quantities of interest in the corresponding tensor network states, by means of a mapping to the evaluation of partition functions of classical statistical models, like generalized Ising models with boundary pinning fields. 

In what follows, we shall reconsider the random state averaging technique for a specific class of large dimensional random tensor networks endowed with extra symmetry constraints.
%When each leg of each tensor in the network has dimension D, these statistical models have inverse temperature $ GG$. For large enough $D$, they are in the long-range ordered phase, and we find that the entropies of a boundary region is directly related to the energy of a domain wall between different domains of the order parameter. The minimal energy condition of this domain wall naturally leads to the Ryu-Takayanagi formula. Besides obtaining the RT formula for multiple intervals, the technique of random state averaging allows us to study many further properties of a random tensor network:

\section{Random Tensor Network States with Gauge Symmetry constraints}\label{const} 

Now let us consider a tensor $T_{\lambda_1\cdots\lambda_v}$ which satisfies the following symmetry 
\begin{eqnarray}\label{eq:sym}
T_{\lambda_1\cdots\lambda_v} = T_{[\lambda_1+\ell]_D\cdots[\lambda_v+\ell]_D}, \quad  \forall\ell \in \mathbb{Z}
\end{eqnarray}
The square bracket $[\cdots]_D$ denotes the modular arithmetic: for all $k\in \mathbb{Z}$ and $D\in\mathbb{Z}^+$
\begin{eqnarray}
[k]_D \equiv k \text{ mod } D, \quad [k]_D \in \mathbb{Z}/\!D
\end{eqnarray}
where $\mathbb{Z}/\!D$ is the the set of integers modulo $D$. Such a tensor state can be considered as a particular case of a GFT tensor field, with arguments/indices taking values on a finite group.%\footnote{In a (tensorial) group field theory, the gauge symmetry defined in \eqref{symme} can also be imposed as a ``dynamical'' feature via the choice of specific kernels in the action.} 

For all $k_1,k_2\in\mathbb{Z}$ and $D\in\mathbb{Z}^+$, it satisfies:
\begin{eqnarray}\label{eq:ruleMod}
&[k_1]_D&+[k_2]_D=\left[k_1+k_2\right]_D, \\ \nonumber
&[[k_1]_D&+[k_2]_D]_D=[k_1+k_2]_D, \quad [D]_D=0
\end{eqnarray}

In the presence of the symmetry \eqref{eq:sym}, only $D^{v-1}$ components are independent. We choose a new gauge 
\begin{eqnarray}\label{eq:gaugenew}
\mu= \lambda_1D^{v-1} + \sum_{a=2}^{v}[\lambda_a-\lambda_1]_D D^{v-a}
\end{eqnarray}
such that
\begin{equation}
T_\mu=\bigoplus_{i=0}^{D-1} T_{\mu_i}\equiv T_{\lambda_1\cdots\lambda_v}
\end{equation}
and numerically 
\begin{eqnarray}
T_{\mu_i} = T_{\mu_j}, \quad \forall i, j = 0,1,\dots,D-1 
\end{eqnarray}
In a different representation, for a given component $T_\mu$ one has
\begin{eqnarray}
T_{\mu}=T_{[\mu+\ell D^{v-1}]_{D^v}}, \quad \forall \ell\in\mathbb{Z}
\end{eqnarray}

%Because of the gauge choice \eqref{eq:gauge}, $T_\mu$ can be written as a direct sum of $D$ vectors
%\begin{eqnarray}
%T_{\mu}\equiv 
%\begin{pmatrix}
%T_{\mu_0}\\
%T_{\mu_1}\\
%\vdots\\
%T_{\mu_{D-1}}
%\end{pmatrix},
%\mu_i = 0,~1,\dots, ~D^{v-1}-1, 
%\end{eqnarray}
%$\forall i=0,1,\dots,D-1$, in which 
%the components in $T_{\mu_i}$ are independent from each other. However one can observe that if we use the gauge choice \Ref{eq:gauge}
%\begin{eqnarray}
%T_{\mu_i}\neq T_{\mu_j}, \quad \forall i\neq j
%\end{eqnarray} 
%In order to avoid this inconsistency, we choose a new gauge $T_\mu=\bigoplus_{i=0}^{D-1} T_{\mu_i}\equiv T_{\lambda_1\cdots\lambda_v}$, with
%\begin{eqnarray}\label{eq:gaugenew}
% \mu= \lambda_1D^{v-1} + \sum_{a=2}^{v}[\lambda_a-\lambda_1]_D D^{v-a}
%\end{eqnarray}
%In the new gauge, numerically the components satisfy
%\begin{eqnarray}
%T_{\mu_i} = T_{\mu_j}, \quad \forall i, j
%\end{eqnarray}
%and in direct summed representation, for a given component $T_\mu$
%\begin{eqnarray}
%T_{\mu}=T_{[\mu+\ell D^{v-1}]_{D^v}}, \quad \forall \ell\in\mathbb{Z}
%\end{eqnarray}
From now on, for simplicity, we denote the tensor with the symmetry \eqref{eq:sym} with two indices: $\mu$ and $ i = \lambda_1\in \mathbb{Z}/\!D$ as $T_{\mu_i} = T_{\lambda_1\cdots\lambda_v}$, with
\begin{eqnarray}
\mu_i = \sum_{a=2}^{v}[\lambda_a-\lambda_1]_D D^{v-a}\in \mathbb{Z}/\!D^{v-1}
\end{eqnarray}

For a given $i$, the vector $T_{\mu_i}$ is lying on a $D^{v-1}$ dimensional space, which is a fundamental representation space of the group $\U(D^{v-1})$. Because $\overline{T}_\mu T_{\mu}=1$ and \eqref{eq:sym}, $T_{\mu_i}$ is also normalized
\begin{eqnarray}
\overline{T}_{\mu_i}T_{\mu_i}=D^{-1},\quad \forall i=0,1,\cdots,D-1
\end{eqnarray}
Then the {\emph random} nature of the tensor $T_{\lambda_1\cdots\lambda_v}$ implies that, with respect to the same $T_0^{\mu}$, the group element $U_n \in \U(D^{v-1})$ is randomly chosen for each node.

\section{entanglement entropy of a RTN subregion}
For the specific class of  RTN states defined in Section  \ref{const}, we shall then investigate the effect of the symmetry constraint on the holographic behaviour of the entanglement entropy. We follow a similar procedure as in \cite{hayden}.

\begin{figure}[t!]
\centering 
\includegraphics[width=.3\textwidth]{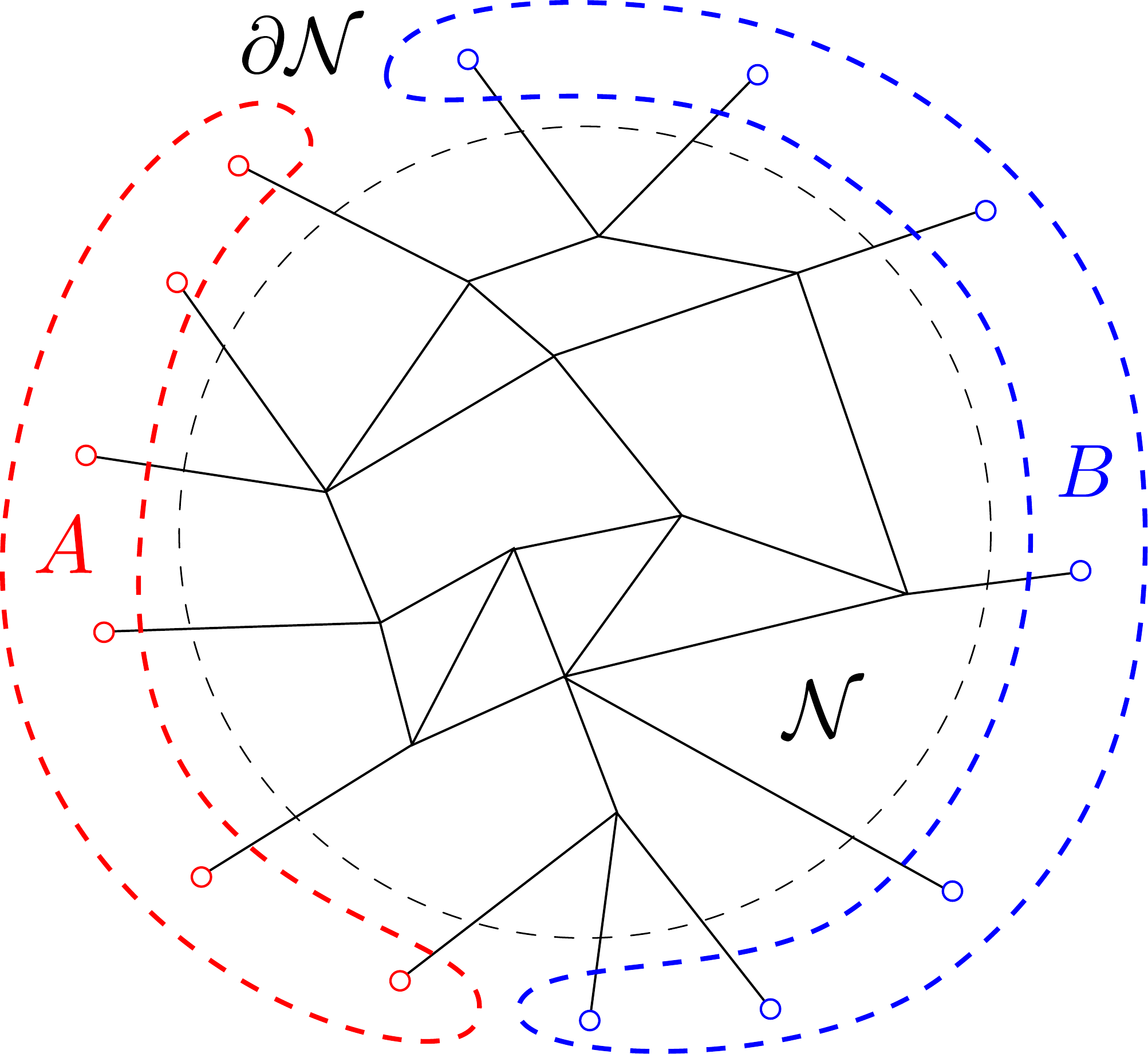}
\caption{\label{fig:AB} Boundary $\p\mathcal{N}$ of network $\mathcal{N}$ divided into two parts $A$ and $B$.}\label{AB}
\end{figure}

Given our tensor network state, 
\begin{equation}
\ket{\Psi_{\mathcal{N}}}\equiv \bigotimes_\ell^L\bra{M_{\ell}}\bigotimes_n^N\ket{T_n} \in \mathbb{H}_{\partial\mathcal{N}},
\end{equation}
with tensor states satisfying the relation given in \eqref{eq:sym}, we consider a bipartition of the boundary Hilbert space
\begin{equation}\label{bipart}
\mathbb{H}_{\partial\mathcal{N}}= \mathbb{H}_{A}\otimes \mathbb{H}_{B}
\end{equation}
associated to the definition of two -- a priori not adjacent -- boundary subregions $A$ and $B$ (see Figure \ref{AB}).
 
A measure of the entanglement between the two subsystems is given by the von Neumann entropy of the reduced density matrix of the subsystem, either $A$ or $B$, defined by partial tracing over the full system  Hilbert space. Focussing on subsystem $A$, for $ \rho \equiv \ket{\Psi_{\mathcal{N}}} \bra{\Psi_{\mathcal{N}}}
$, we have
\begin{eqnarray}
\rho_A = \text{Tr}_B (\rho) , 
\end{eqnarray}
and the entanglement entropy between $A$ and $B$ is given by the von Neumann entropy
\begin{eqnarray}
S_{\text{EE}}(A)=-\text{Tr} [\, \widehat{\rho}_A \ln \widehat{\rho}_A\,] , 
\end{eqnarray} 
where now
\begin{eqnarray}
\widehat{\rho}_A \equiv \frac{\rho_A}{\text{Tr}\rho}
\end{eqnarray}
is the normalized reduced density matrix.

Because of the difficulties in computing directly the von Neumann entropy,  one uses the so-called replica trick to approach the problem. Contracting $N$ copies of the reduced density matrix $\rho_A$ and taking the logarithm of the trace of $\rho_A^N$, one obtains the $N$th-order R\'{e}nyi entropy
\begin{eqnarray}\label{eq:Sn}
S_N(A)=-\frac{1}{N-1}\ln\text{Tr}\widehat{\rho}_A^{\,N} \qquad .
\end{eqnarray}
We can also define
\begin{eqnarray}\label{eq:ZnZ0}
Z_A^{(N)} & \equiv & \text{Tr}\rho_A^N = \text{Tr} [\rho^{\otimes N} \mathds{P}(\pi^0_A;N,d)], \\ 
\nonumber Z_0^{N} & \equiv & (\text{Tr}\rho)^N, 
\end{eqnarray}
where $\mathds{P}(\pi^0_A;N,d)$ is the 1-cycle permutation operator acting on the states in region $A$,
\begin{eqnarray}
\mathds{P}(\pi^0_A;N,d)=\prod_{s=1}^N\delta_{\mu_A^{([s+1]_N)}\mu_A^{(s)}}
\end{eqnarray}
and $d$ is the dimension of the Hilbert space in the same region $A$.

The replica trick is useful because the R\'{e}nyi entropy $S_N$, which is easier to compute, coincides with the von Neumann entropy of region $A$, and thus with the entanglement entropy between regions $A$ and $B$, in the limit $N \rightarrow 1$
\begin{eqnarray}
S_{\text{EE}}(A)=\lim_{N\rightarrow 1}S_N(A)
\end{eqnarray}

\subsection{Derivation of the Second R\'enyi Entropy}
As a first step, let us write down the second order R\'enyi entropy, $S_2$, for a generic tensor network state $\ket{\Psi_\mathcal{N}}$ in $\mathbb{H}_{\p\mathcal{N}}$. We rewrite states in the following synthetic index notation,
\begin{eqnarray}
\ket{\Psi_\mathcal{N}} &\Longleftrightarrow & \Psi_{\{\lambda_A\}\{\lambda_B\}}\equiv \Psi_{AB} \\
\bigotimes_n\ket{T_n} &\Longleftrightarrow & \left(\bigotimes_nT_n\right)_{\{\lambda_A\}\{\lambda_B\}\{\lambda_C\}}\equiv \mathcal{T}_{ABC} \\
\bigotimes_\ell\bra{M_{\ell}} &\Longleftrightarrow & \left(\bigotimes_\ell \overline{M}_\ell\right)_{\{\lambda_C\}}\equiv \overline{\mathcal{M}}_C
\end{eqnarray}
Based on the definition $\Ref{eq:TNS}$, the tensor network state is rewritten as
\begin{eqnarray}
\Psi_{AB}= \overline{\mathcal{M}}_C\mathcal{T}_{ABC}
\end{eqnarray}
where $A$ and $B$ label the two regions of the boundary $\p\mathcal{N}$.

All internal links are contracted with nodes. The density matrix corresponding to $\Psi_{AB}$ is
\begin{eqnarray}
\rho_{\overline{A}A\overline{B}B}= \overline{\Psi}_{\overline{A}\overline{B}}\Psi_{AB}
\end{eqnarray}
Then $S_2$ is defined as
\begin{eqnarray}
\ex^{-S_2}=\frac{Z_A^{(2)}}{Z_0^{(2)}}
\end{eqnarray}
with $Z_0^{(2)}= \tr[\rho^{\otimes 2}]$ and $Z_A^{(2)} = \tr[\rho_A^{\otimes 2}]$.

In particular, for $N=2$, the cyclic group $\mathcal{S}_n$ only has two elements: the identity $\mathds{1}$ and swap operator $\mathds{F}$, so that
\begin{eqnarray}
\mathds{P}(\pi^0_A;2,d) \equiv \mathds{F}^{(A)}
\end{eqnarray}
Then
\begin{eqnarray}
Z_A^{(2)} &=& \tr[\rho_A^{\otimes 2}] = \tr[\rho^{\otimes 2}\mathds{F}^{(A)}]\\ \nonumber
&=&\rho_{\overline{A}_1A_1\overline{B}_1B_1}\rho_{\overline{A}_2A_2\overline{B}_2B_2}\mathds{F}^{(A)}_{\overline{A}_1A_1\overline{A}_2A_2}\mathds{1}^{(B)}_{\overline{B}_1B_1\overline{B}_2B_2}
\end{eqnarray}
\begin{eqnarray}
Z_0^{(2)}&=& \tr[\rho^{\otimes 2}]=\\ \nonumber
&=&\rho_{\overline{A}_1A_1\overline{B}_1B_1}\rho_{\overline{A}_2A_2\overline{B}_2B_2}\mathds{1}^{(A)}_{\overline{A}_1A_1\overline{A}_2A_2}\mathds{1}^{(B)}_{\overline{B}_1B_1\overline{B}_2B_2}
\end{eqnarray}
Following ~\cite{hayden}, given the random nature of the tensor network, one can consider the average of the relevant quantities over all states in the single-vertex Hilbert space and look for the typical value of the entropy. 

The entropy average can be expanded in powers of the fluctuations $\delta Z_A^{(2)} = Z_A^{(2)} - \mathbb{E}({Z_A^{(2)}})$ and $\delta Z_0^{(2)} = Z_0^{(2)} - \mathbb{E}({Z_0^{(2)}})$, so that \cite{hayden}
\begin{eqnarray}
\mathbb{E}({S_N(A)})&=&-\mathbb{E}\left({{\log\frac{\mathbb{E}({Z_A^{(2)}})+\delta  Z_A^{(2)}}{\mathbb{E}({Z_0^{(2)}})+\delta Z_0}}}\right)\\ \nonumber
&=&-\log \frac{\mathbb{E}({Z_A^{(2)}})}{\mathbb{E}({Z_0^{(2)}})}+\text{fluctuations}
\end{eqnarray}
In large enough bond dimensions $D$, as a direct consequence of the \emph{concentration of measure phenomenon} \cite{hayden2}, the statistical fluctuations around the average value are exponentially suppressed. Therefore, it is possible to approximate the entropy with high probability by the averages of $Z_A^{(2)}$ and $Z_0^{(2)} $,
\begin{equation}
\mathbb{E}({S_N(A)})\simeq-\log \frac{\mathbb{E}({Z_A^{(2)}})}{\mathbb{E}({Z_0^{(2)}})}
\end{equation}
In order to get the typical R\'enyi entropy one needs then to compute $\mathbb{E}({Z_A^{(2)}})$ and $\mathbb{E}({Z_0^{(2)}})$ separately.

\subsection{Random State Averaging}

Let us first consider the case without the gauge symmetry \Ref{eq:sym} for a given graph with only one node. The corresponding density matrix is
\begin{equation}
\rho_{\mu\overline{\mu}}\equiv T_{\mu}\overline{T}_{\overline{\mu}}
\end{equation}
Consider $N$ copies of the density matrix $\rho^{\otimes N}$. If $T$ is uniformly distributed, then the average $\rho^{\otimes N}$ over $T$ is given by
\begin{equation}
\mathbb{E}_T (\rho^{\otimes N}) \equiv \int_{\U(D^v)} \dd U ~ \prod_{s=1}^N (UT^0)_{\mu^{(s)}}\overline{(UT^0)}_{\bar{\mu}^{(s)}}
\end{equation}
Because of the Schur's lemma, since $\mathbb{H}_{D^v}$ is the irrep of $\U(D^v)$, the result of the average is the identity matrix on the symmetric subspace of $\mathbb{H}_{D^v}^{\otimes n}$. When $\tr\rho=1$, the result is then
\begin{eqnarray}
\mathbb{E}_T (\rho^{\otimes N })&=&\frac{1}{D^v[N]}\sum_{\pi\in \mathcal{S}_N}\prod_{s=1}^N\delta_{\mu^{(s)}\overline{\mu}^{(\pi(s))}}\\ \nonumber
&\equiv& \frac{1}{D^v[N]}\sum_{\pi\in \mathcal{S}_N}\mathds{P}(\pi;N,D^v)
\end{eqnarray}
where
\begin{equation}\label{eq:dimPsym}
D^v[N]\equiv D^v(D^v+1)\cdots(D^v+N-1)\quad.
\end{equation}

$\mathcal{S}_N$ is the symmetric group on $N$ objects and 
\begin{equation}
\mathds{P}(\pi;N,D^v)\equiv \prod_{s=1}^N\delta_{\mu^{(s)}\overline{\mu}^{(\pi(s))}}
\end{equation}
which is the representation matrix of $\pi\in\mathcal{S}_N$ on $\mathbb{H}_{D^v}^{\otimes N}$. Using the gauge \Ref{eq:gauge} or \Ref{eq:gaugenew}, we can get the relation between representations on $\mathbb{H}_{D^v}$ and $\mathbb{H}_{D}^{\otimes v}$
\begin{equation}
\delta_{\mu^{(s)}\overline{\mu}^{(\pi(s))}}\equiv \prod_{a=1}^v\delta_{\lambda_a^{(s)}\overline{\lambda}_{a}^{(\pi(s))}}
\end{equation}
Then we have 
\begin{equation}
\mathds{P}(\pi;N,D^v) = \prod_{a=1}^v \mathds{P}(\pi_a;N,D)
\end{equation}

If $T$ is an random Gaussian vector, then the average is
\begin{equation}
\mathbb{E}_T (\rho^{\otimes N}) \equiv \int \mathcal{D}T~ \ex^{-\beta |T|^2}\prod_{s=1}^N T_{\mu^{(s)}}\overline{T}_{\overline{\mu}^{(s)}}
\end{equation}
If we ask $\mathbb{E}_T (\rho)=\mathds{1}/D^v$ and $T=x \hat{T}$, where $|\hat{T}|=1$, then $\beta=D^v$ and the average $\mathbb{E}_T (\rho^{\otimes N}) $ becomes
\begin{eqnarray}
&\, & \int \dd x ~|x|^{2N} \ex^{-D^v |x|^2}\int\dd U~ \prod_{s=1}^N (U\hat{T}^0)_{\mu^{(s)}}\overline{(U\hat{T}^0)}_{\overline{\mu}^{(s)}}\nonumber\\
&\qquad&= (D^v)^{-N}\sum_{\pi\in \mathcal{S}_N}\prod_{s=1}^N\delta_{\mu^{(s)}\overline{\mu}^{(\pi(s))}}
\end{eqnarray}

Now let us consider the case with symmetry \Ref{eq:sym}. The corresponding density matrix is
\begin{equation}
\rho_{\mu\mu'}\equiv T_{\mu}\overline{T}_{\overline{\mu}}=T_{\mu_i}\overline{T}_{\overline{\mu}_{\overline{i}}}\equiv  \widetilde{\rho}_{\mu_i\overline{\mu}_{\overline{i}}}
\end{equation}
The expression for the $N$ copies of the density matrix reads
\begin{equation}
\rho^{\otimes N}= \prod_{s=1}^N\widetilde{\rho}_{\mu_{i(s)}\overline{\mu}_{\overline{i}(s)}}=\prod_{s=1}^N T_{\mu_{i(s)}}\overline{T}_{\overline{\mu}_{\overline{i}(s)}}
\end{equation}

If $T$ is uniformly distributed, then the average of $\rho^{\otimes N}$ over $T$ is
\begin{equation}
\mathbb{E}_T (\rho^{\otimes N})=\int_{\U(D^{v-1})}\dd U \prod_{s=1}^N (UT^0)_{\mu_{i(s)}}\overline{(UT^0)}_{\overline{\mu}_{\overline{i}(s)}}
\end{equation}
As shown in the first case, the result of the integral is the identity matrix in the symmetric subspace of $\mathbb{H}_{D^{v-1}}^{\otimes N}$. 
\begin{eqnarray}\label{eq:ETS}
\mathbb{E}_T (\rho^{\otimes N})&\propto&\sum_{\pi\in\mathcal{S}_N}\prod_{s=1}^N\delta_{\mu_{i(s)}\overline{\mu}_{\overline{i}(\pi(s))}}\\ \nonumber
&\equiv& \sum_{\pi\in\mathcal{S}_N}\mathds{P}_{\{i(s)\}\{\overline{i}(s)\}}(\pi;N,D^{v-1})
\end{eqnarray}
%The normalization will be calculated in the next subsection. 
where $\mathds{P}_{\{i(s)\}\{\overline{i}(s)\}}(\pi;N,D^{v-1})$ is the representation matrix of $\pi\in\mathcal{S}_N$ on $\mathbb{H}_{D^{v-1}}^{\otimes N}$ with a set of ${\{i(s)\}\{\overline{i}(s)\}}$. Similarly, when $T_{\mu}$ is a Gaussian vector, the result of the average is the same as the above equation up to a normalization. The details of the matrix in \eqref{eq:ETS} are given in Appendix \ref{Imatrix}.

By using the gauge \Ref{eq:gaugenew}, one can show the relation between the representations $\mu_i$ and $\lambda_a$. Because of \Ref{eq:gaugenew}, $\delta_{\mu_{i(s)}\overline{\mu}_{\overline{i}(s')}}$ is not zero only when
\begin{equation}
[\lambda_a(s)-i(s)]_D=[\overline{\lambda}_a(s')-\overline{i}(s')]_D
\end{equation} 
because of the modular rules \Ref{eq:ruleMod}, the above equation can be rewritten as
\begin{equation}
[\lambda_a(s)-\overline{\lambda}_a(s')]_D=[i(s)-\overline{i}(s')]_D
\end{equation}
If $[i(s)-\overline{i}(s')]_D=\ell\in\mathbb{Z}/\!D$, then
\begin{equation}
\delta_{\mu_{i(s)}\overline{\mu}_{\overline{i}(s')}}=\prod_{a=1}^v \delta_{[\lambda_{a}(s)-\overline{\lambda}_a(s')]_D,\ell}= \prod_{a=1}^v \delta_{[\lambda_{a}(s)-\ell]_D,\overline{\lambda}_a(s')} \nonumber
\end{equation}
Notice that $\ell$ is a uniform shift for all legs of each node as long as the step $[i(s)-\overline{i}(s')]_D=\ell$ is fixed. 
Finally, we define for later use the trace on a tensor $\mathbb{T}_{\{\mu^{(s)}\}\{\overline{\mu}^{(s)}\}}$, as
\begin{equation}
\tr{\mathbb{T}}=\mathbb{T}_{\{\mu^{(s)}\}\{\overline{\mu}^{(s)}\}}\prod_{s=1}^N\delta_{\mu^{(s)}\overline{\mu}^{(s)}} \quad ,
\end{equation}
which becomes, with the symmetry \Ref{eq:sym},
\begin{equation}
\tr{\mathbb{T}}=\sum_{\substack{\{i(s)\}\\ \{\overline{i}(s)\}}}\mathbb{T}_{\{\mu_{i(s)}\}\{\overline{\mu}_{\overline{i}(s)}\}}\prod_{s=1}^N\delta_{\mu_{i(s)}\overline{\mu}_{\overline{i}(s)}}\delta_{i(s)\overline{i}(s)} \quad .
\end{equation}
\,\\

\subsection{$S_2$ R\'enyi with symmetry constraint}
Coming back to the case of $N=2$, we can now explicitly write down the expression for the average of the single symmetric tensor defined in \Ref{eq:sym},
\begin{eqnarray}
\mathbb{E}_T (\rho_{T_n}^{\otimes 2})= \frac{1}{D^2 D^{v-1}[2]}\sum_{\footnotesize{\substack{m(1)\\m(2)}}}\left(\mathds{1}_{m(1)m(2)}+\mathds{F}_{m(1)m(2)}\right) \qquad
\end{eqnarray}
Given \Ref{eq:dimPsym}, $D^{v-1}[2] = D^{v-1}(D^{v-1}+1)$, we can define the normalization as
$\mathbb{D}_2=D^2 D^{v-1}[2]$.
Therefore, for the density matrix of a tensor network with $N$ nodes we have
\begin{widetext}
\begin{eqnarray}\label{ave}
\mathbb{E}_T(\rho^{\otimes 2})&=& \tr_C\left[\bigotimes_n \mathbb{E}_T (\rho_{T_n}^{\otimes 2})~\rho_M^{\otimes 2}\right]= \frac{1}{\mathbb{D}_2^N}\tr_C\left[\bigotimes_n^N \sum_{m_n(1)m_n(2)}\left(\mathds{1}_{m_n(1)m_n(2)}+\mathds{F}_{m_n(1)m_n(2)}\right)~\rho_M^{\otimes 2}\right]\nonumber\\
&=& \frac{1}{\mathbb{D}_2^N}\sum_{\{m_n(1)\}\{m_n(2)\}}\tr_C\left[\bigotimes_n\mathds{1}_{m_n(1)m_n(2)} \bigotimes_{n'}\mathds{F}_{m_{n'}(1)m_{n'}(2)}~\rho_M^{\otimes 2}\right]
\end{eqnarray}
\end{widetext}

\noindent
In the case of a network of $N$-nodes, the above sum is naively given by a sum of $(2D^2)^N$ terms, $2D^2$ choices for each node, but with several terms vanishing. 

In order to calculate \eqref{ave} explicitly, it is convenient to separately consider the case of internal and boundary links. For an internal links connecting two nodes, we have the following three cases: 
\begin{itemize}
\item $\mathds{1}_{m(1)m(2)}$ and $\mathds{1}_{m'(1)m'(2)}$
\begin{eqnarray}
&&\tr\left[\mathds{1}_{m(1)m(2)}~\rho_{M_{\ell}}^{\otimes 2}~\mathds{1}_{m'(1)m'(2)}\right]\nonumber\\
%&=& \delta_{[\lambda(1)-m(1)]_D,\overline{\lambda}(1)}\delta_{[\lambda(2)-m(2)]_D,\overline{\lambda}(2)}\delta_{[\lambda'(1)-m'(1)]_D,\overline{\lambda}'(1)}\times\nonumber\\
%&&\times \delta_{[\lambda'(2)-m'(2)]_D,\overline{\lambda}'(2)}
%\delta_{\lambda(1)\lambda'(1)}\delta_{\overline{\lambda}(1),\overline{\lambda}'(1)}\delta_{\lambda(2)\lambda'(2)}\delta_{\overline{\lambda}(2),\overline{\lambda}'(2)}\nonumber\\
%&=&\delta_{[\lambda(1)-m(1)]_D,\overline{\lambda}(1)}\delta_{[\lambda(2)-m(2)]_D,\overline{\lambda}(2)}\delta_{[\lambda(1)-m'(1)]_D,\overline{\lambda}(1)} \nonumber \\
%&&\times \delta_{[\lambda(2)-m'(2)]_D,\overline{\lambda}(2)}\nonumber\\
%&=&\delta_{[\lambda(1)-m(1)]_D,[\lambda(1)-m'(1)]_D}\delta_{[\lambda(2)-m(2)]_D,[\lambda(2)-m'(2)]_D}\nonumber\\
&=&D^2~\delta_{m(1)m'(1)}\delta_{m(2)m'(2)}
\end{eqnarray}
\item $\mathds{F}_{m(1)m(2)}$ and $\mathds{F}_{m'(1)m'(2)}$
\begin{eqnarray}
&&\tr\left[\mathds{F}_{m(1)m(2)}~\rho_{M_{\ell}}^{\otimes 2}~\mathds{F}_{m'(1)m'(2)}\right]\nonumber\\
%&=& \delta_{[\lambda(1)-m(1)]_D,\overline{\lambda}(2)}\delta_{[\lambda(2)-m(2)]_D,\overline{\lambda}(1)}\delta_{[\lambda'(1)-m'(1)]_D,\overline{\lambda}'(2)}\times\nonumber\\
%&&\times \delta_{[\lambda'(2)-m'(2)]_D,\overline{\lambda}'(1)}
%\delta_{\lambda(1)\lambda'(1)}\delta_{\overline{\lambda}(1),\overline{\lambda}'(1)}\delta_{\lambda(2)\lambda'(2)}\delta_{\overline{\lambda}(2),\overline{\lambda}'(2)}\nonumber\\
%&=&\delta_{[\lambda(1)-m(1)]_D,\overline{\lambda}(2)}\delta_{[\lambda(2)-m(2)]_D,\overline{\lambda}(1)}\delta_{[\lambda(1)-m'(1)]_D,\overline{\lambda}(2)}\times \nonumber\\
%&&\times \delta_{[\lambda(2)-m'(2)]_D,\overline{\lambda}(1)}\nonumber\\
%&=&\delta_{[\lambda(1)-m(1)]_D,[\lambda(1)-m'(1)]_D}\delta_{[\lambda(2)-m(2)]_D,[\lambda(2)-m'(2)]_D}\nonumber\\
&=&D^2~\delta_{m(1)m'(1)}\delta_{m(2)m'(2)}
\end{eqnarray}
\item $\mathds{1}_{m(1)m(2)}$ and $\mathds{F}_{m'(1)m'(2)}$
\begin{eqnarray}
&&\tr\left[\mathds{1}_{m(1)m(2)}~\rho_{M_{\ell}}^{\otimes 2}~\mathds{F}_{m'(1)m'(2)}\right]\nonumber\\
%&=& \delta_{[\lambda(1)-m(1)]_D,\overline{\lambda}(1)}\delta_{[\lambda(2)-m(2)]_D,\overline{\lambda}(2)}\delta_{[\lambda'(1)-m'(1)]_D,\overline{\lambda}'(2)}\times\nonumber\\
%&&\times\delta_{[\lambda'(2)-m'(2)]_D,\overline{\lambda}'(1)}\delta_{\lambda(1)\lambda'(1)}\delta_{\overline{\lambda}(1),\overline{\lambda}'(1)}\delta_{\lambda(2)\lambda'(2)}\delta_{\overline{\lambda}(2),\overline{\lambda}'(2)}\nonumber\\
%&=&\delta_{[\lambda(1)-m(1)]_D,\overline{\lambda}(1)}\delta_{[\lambda(2)-m(2)]_D,\overline{\lambda}(2)}\delta_{[\lambda(1)-m'(1)]_D,\overline{\lambda}(2)} \nonumber \\
%&& \delta_{[\lambda(2)-m'(2)]_D,\overline{\lambda}(1)}\nonumber\\
%&=&\delta_{[\lambda(1)-m(1)]_D,[\lambda(2)-m'(2)]_D}  \delta_{[\lambda(2)-m(2)]_D,[\lambda(1)-m'(1)]_D}\nonumber\\
%&=&\delta_{[\lambda(1)-m(1)+m'(2)]_D,[\lambda(1)-m'(1)+m(2)]_D} \nonumber \\
&=&D~\delta_{[m(1)+m(2)]_D,[m'(1)+m'(2)]_D}
\end{eqnarray}
\end{itemize}

On the boundary of region $A$, for $\mathds{F}^{(A)}= \mathds{F}_{00}$ at one end of the boundary link, we have instead only two cases:
\begin{itemize}
\item $\mathds{F}^{(A)}$ and $\mathds{F}_{m'(1)m'(2)}$
\begin{eqnarray}
\tr\left[\mathds{F}^{(A)}~\rho_{M_{\ell}}^{\otimes 2}~\mathds{F}_{m'(1)m'(2)}\right]&=&\tr \nonumber\left[\mathds{F}_{00}~\rho_{M_{\ell}}^{\otimes 2}~\mathds{F}_{m'(1)m'(2)}\right]\\ 
&=&D^2~\delta_{0 m'(1)}\delta_{0m'(2)}
\end{eqnarray}
\item $\mathds{F}^{(A)}$ and $\mathds{1}_{m'(1)m'(2)}$
\begin{eqnarray}
\tr\left[\mathds{F}^{(A)}~\rho_{M_{\ell}}^{\otimes 2}~\mathds{1}_{m'(1)m'(2)}\right]&=&\tr \nonumber\left[\mathds{F}_{00}~\rho_{M_{\ell}}^{\otimes 2}~\mathds{1}_{m'(1)m'(2)}\right] \\ 
&=&D~\delta_{0,[m'(1)+m'(2)]_D}
\end{eqnarray}
\end{itemize}

Finally, on the boundary of region $B$, where $\mathds{1}^{(B)}=\mathds{1}_{00}$ at one end of the boundary link, there exist two cases,
\begin{itemize}
\item $\mathds{1}^{(B)}$ and $\mathds{F}_{m'(1)m'(2)}$
\begin{eqnarray}
\tr\left[\mathds{1}^{(B)}~\rho_{M_{\ell}}^{\otimes 2}~\mathds{F}_{m'(1)m'(2)}\right]&=&\tr \nonumber\left[\mathds{1}_{00}~\rho_{M_{\ell}}^{\otimes 2}~\mathds{F}_{m'(1)m'(2)}\right]\\ 
&=&D~\delta_{0,[m'(1)+m'(2)]_D}
\end{eqnarray}
\item $\mathds{1}^{(B)}$ and $\mathds{1}_{m'(1)m'(2)}$
\begin{eqnarray}
\tr\left[\mathds{1}^{(B)}~\rho_{M_{\ell}}^{\otimes 2}~\mathds{1}_{m'(1)m'(2)}\right]&=&\tr \nonumber\left[\mathds{1}_{00}~\rho_{M_{\ell}}^{\otimes 2}~\mathds{1}_{m'(1)m'(2)}\right]\\ 
&=&D^2~\delta_{0 m'(1)}\delta_{0m'(2)}
\end{eqnarray}
\end{itemize}
\begin{figure}[t!]\label{fig:ABF1}
\centering 
\includegraphics[width=.3\textwidth]{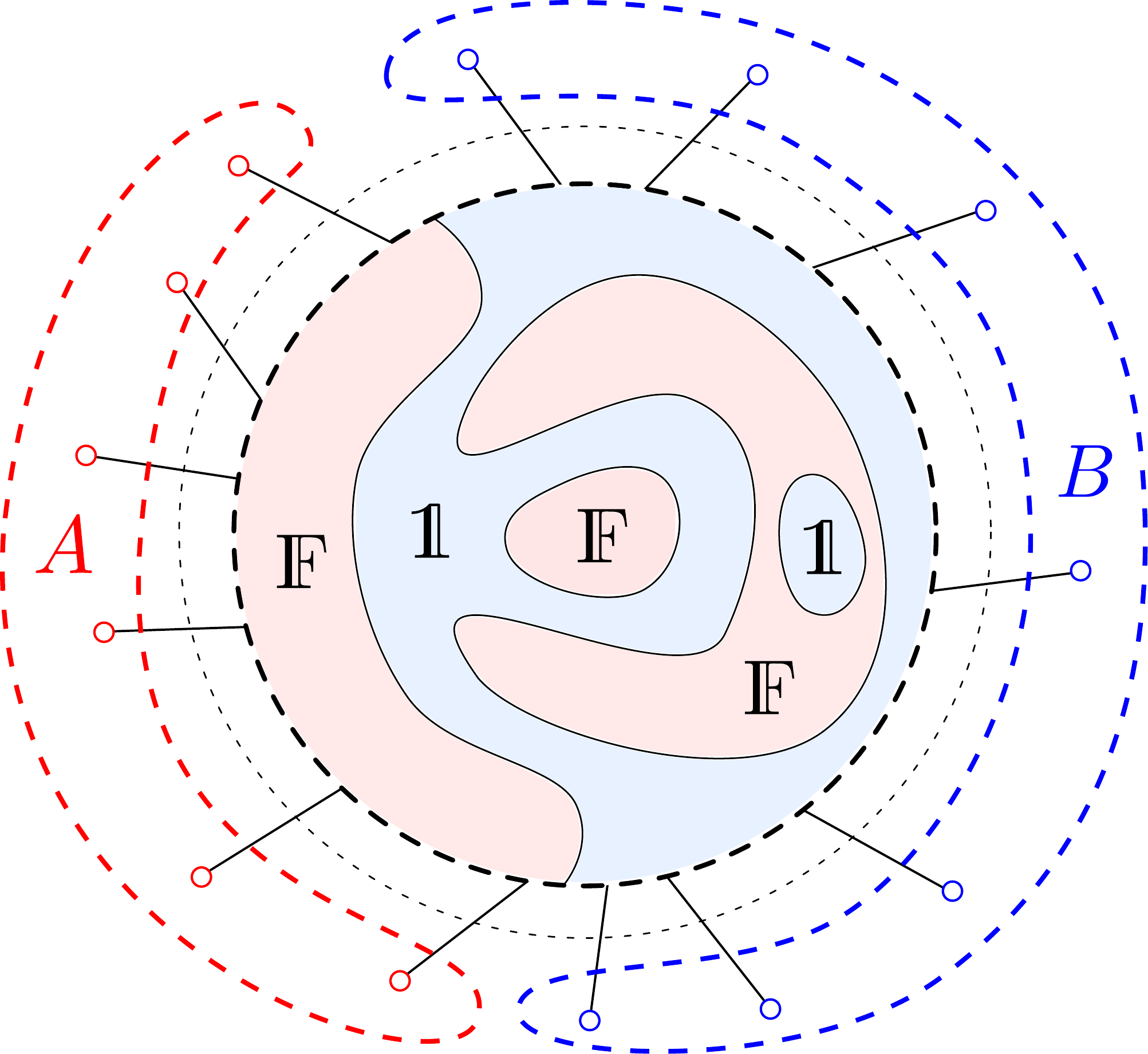}
\caption{Network with boundary condition: $A=\mathds{F}^{(A)}$ and $B=\mathds{1}^{(B)}$ is divided into regions. The nodes in red regions are assigned with $\mathds{F}$ and the ones in blue are with $\mathds{1}$.}
\end{figure} 
\subsection{Remarks on the calculation}
Instead of giving all details about the calculation of the terms identified above, it is enough to sketch the main features of the same.

We see from the results of the previous section that the averaging $Z^{(2)}_A$ and $Z^{(2)}_0$ over $T$ is equivalent to defining a class of networks where each node is assigned a matrix $\mathds{1}_{m(1)m(2)}$ or a matrix $\mathds{F}_{m(1)m(2)}$, and the boundary is assigned $\mathds{F}^{(A)}$, and $\mathds{1}^{(B)}$ for $Z^{(2)}_A$, and $\mathds{1}^{(A)}$ and $\mathds{1}^{(B)}$ for $Z^{(2)}_0$.

For all $\mathds{1}_{m(1)m(2)}$ and $\mathds{F}_{m(1)m(2)}$, the ones with $[m(1)+m(2)]_D\neq 0$ will never contribute to $Z^{(2)}_A$ and $Z^{(2)}_0$. In fact, if there is one node has $[m(1)+m(2)]_D\neq 0$, it will make its neighbouring node satisfying $[m(1)+m(2)]_D\neq 0$, and these nodes will make their neighbouring nodes satisfying $[m(1)+m(2)]_D\neq 0$. Because all nodes connect to the boundary through a certain number of links, the consequence is that the boundary should be $[m(1)+m(2)]_D\neq 0$, but we have assumed that the boundary is assigned by $\mathds{1}_{00}$ or $\mathds{F}_{00}$, i.e. $[m(1)+m(2)]_D = 0$. Therefore, none of the matrices at each node can satisfy $[m(1)+m(2)]_D\neq 0$. So in the following discussion we only consider the matrices of $\mathds{F}_{m(1)m(2)}$ and $\mathds{1}_{m(1)m(2)}$ with $[m(1)+m(2)]_D = 0$

If a node is $\mathds{F}_{m(1)m(2)}$, then its neighbouring nodes can only be $\mathds{F}_{m'(1)m'(2)}$ or $\mathds{1}_{m'(1)m'(2)}$ with $[m'(1)+m'(2)]_D=0$. So the network ends up being divided into several regions, where all nodes are associated with the same matrix. If a region is associated with $\mathds{F}_{m(1)m(2)}$, its neighbouring region can only be with $\mathds{1}_{m'(1)m'(2)}$, and vice versa. An example is shown in Figure \Ref{fig:ABF1}. Each regions a labeled with $\mathds{F}$ or $\mathds{1}$. The boundaries of these regions are called domain walls \cite{}. The domain walls are also in correspondence with links, such that one end of each is assigned with $\mathds{F}$ and the other end with $\mathds{1}$.

As shown in figure \ref{fig:ABF1}, such domain walls form different \emph{patterns} $\mathcal{P}$ for the network. For a given pattern, changing a region's label from $\mathds{F}_{m(1)m(2)}$ to $\mathds{F}_{m'(1)m'(2)}$ will not change the value of its corresponding term in $Z_A^{(2)}$ or $Z_0^{(2)}$. This implies a degeneracy. The degeneracy of the region that does not connect to the boundary is $D$, which is the number of the possible choice of the pair $(m(1),m(2))$. Therefore we have
\begin{eqnarray}
\mathbb{E}_T(Z_A^{(2)})&=&\frac{1}{\mathbb{D}_2^N}\sum_{\mathcal{P}_A}d_{\mathcal{P}_A}Z_{\mathcal{P}_A}^{(2)}, \\ 
\mathbb{E}_T(Z_0^{(2)})&=&\frac{1}{\mathbb{D}_2^N}\sum_{\mathcal{P}_0}d_{\mathcal{P}_0}Z_{\mathcal{P}_0}^{(2)}
\end{eqnarray}
where $d_{\mathcal{P}}$ is the degeneracy of the pattern, which is the product of the degeneracies of each region in this pattern. $Z_{\mathcal{P}}^{(2)}$ is given as
\begin{eqnarray}
Z_{\mathcal{P}}^{(2)} = D^{2(L-L_{\mathcal{P}})}D^{L_{\mathcal{P}}}=D^{2L-L_{\mathcal{P}}}
\end{eqnarray}
where $L$ is the total link number in a given network $\mathcal{N}$, including links across $\p\mathcal{N}$; $L_{\mathcal{P}}$ is the links across the domain walls in $\mathcal{P}$.
\subsection{Holographic behaviour and Ryu-Takanayagi formula}
The main contribution of $\mathbb{E}_T(Z_A^{(2)})$ is the pattern with the least number of links through the domain walls. We call this domain wall with the least number of links the \emph{minimal surface}. One can show that this is true even after the degeneracy $d_{\mathcal{P}}$ is taken into account. In fact, all patterns can be generated from the one only with minimal surface by deforming the minimal surface or adding new regions. Starting from the pattern corresponding to a minimal surface, any deformation will lead to a surface which is not minimal. When adding a region, on the other hand, this will contribute to $d_{\mathcal{P}}$ with $D$ but to the number of domain wall links at least $v>1$, i.e. the valence of a single node, thus in total one has to consider the product $D^{1-v}<1$ to the original $Z_{\mathcal{P}}^{(2)}$. This gives a contribution that is smaller than the original one. So the main contribution to $\mathbb{E}_T(Z_A^{(2)})$ comes from the pattern with only the minimal surface. And in this pattern there are only two regions, which are labeled by $\mathds{F}^{(A)}$ and $\mathds{1}^{(B)}$, respectively:

\begin{eqnarray}
\mathbb{E}_T(Z_A^{(2)})=\frac{1}{\mathbb{D}_2^N}D^{2L-L_{\text{min}}}\left(1+O(D^{-1})\right)
\end{eqnarray}
\begin{figure}[t!]\label{fig:min}
\centering 
\includegraphics[width=.3\textwidth]{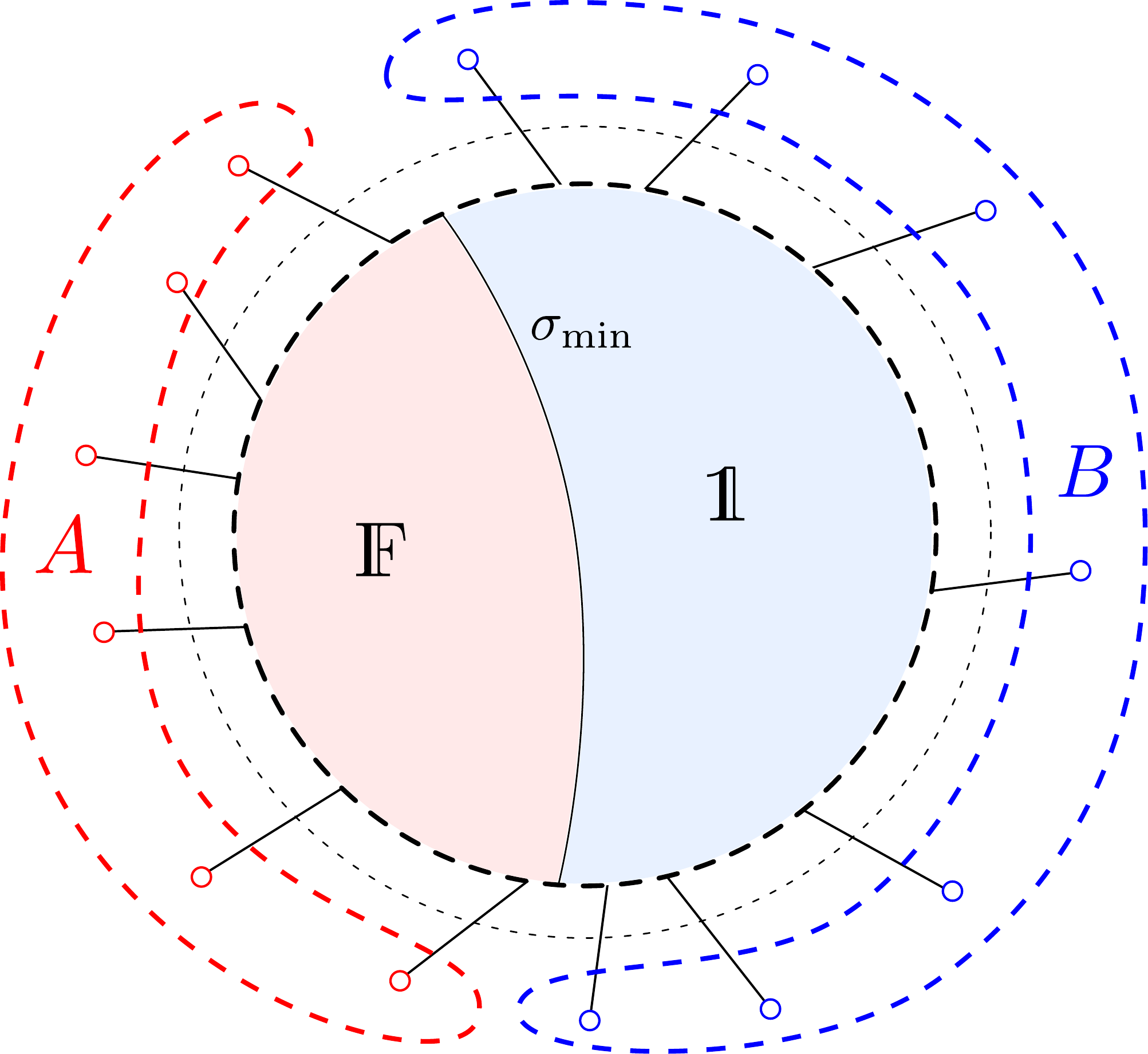}
\caption{The pattern with only the minimal surface $\sigma_{\text{min}}$. Boundary condition: $A=\mathds{F}^{(A)}$ and $B=\mathds{1}^{(B)}$.}
\end{figure} 

In this sense, in agreement with the results in \cite{Hayden}, we find that the main contribution of $\mathbb{E}_T(Z_0^{(2)})$ is given by the pattern without any domain wall. This is because its boundary condition is $\mathds{1}^{(A)}=\mathds{1}^{(B)}=\mathds{1}_{00}$. Such pattern exists and all its nodes are assigned $\mathds{1}_{00}$. Then
\begin{eqnarray}
\mathbb{E}_T(Z_0^{(2)})=\frac{1}{\mathbb{D}_2^N}D^{2L}\left(1+O(D^{-1})\right)
\end{eqnarray}
\begin{figure}[htbp!]\label{fig:Z0}
\centering 
\includegraphics[width=.3\textwidth]{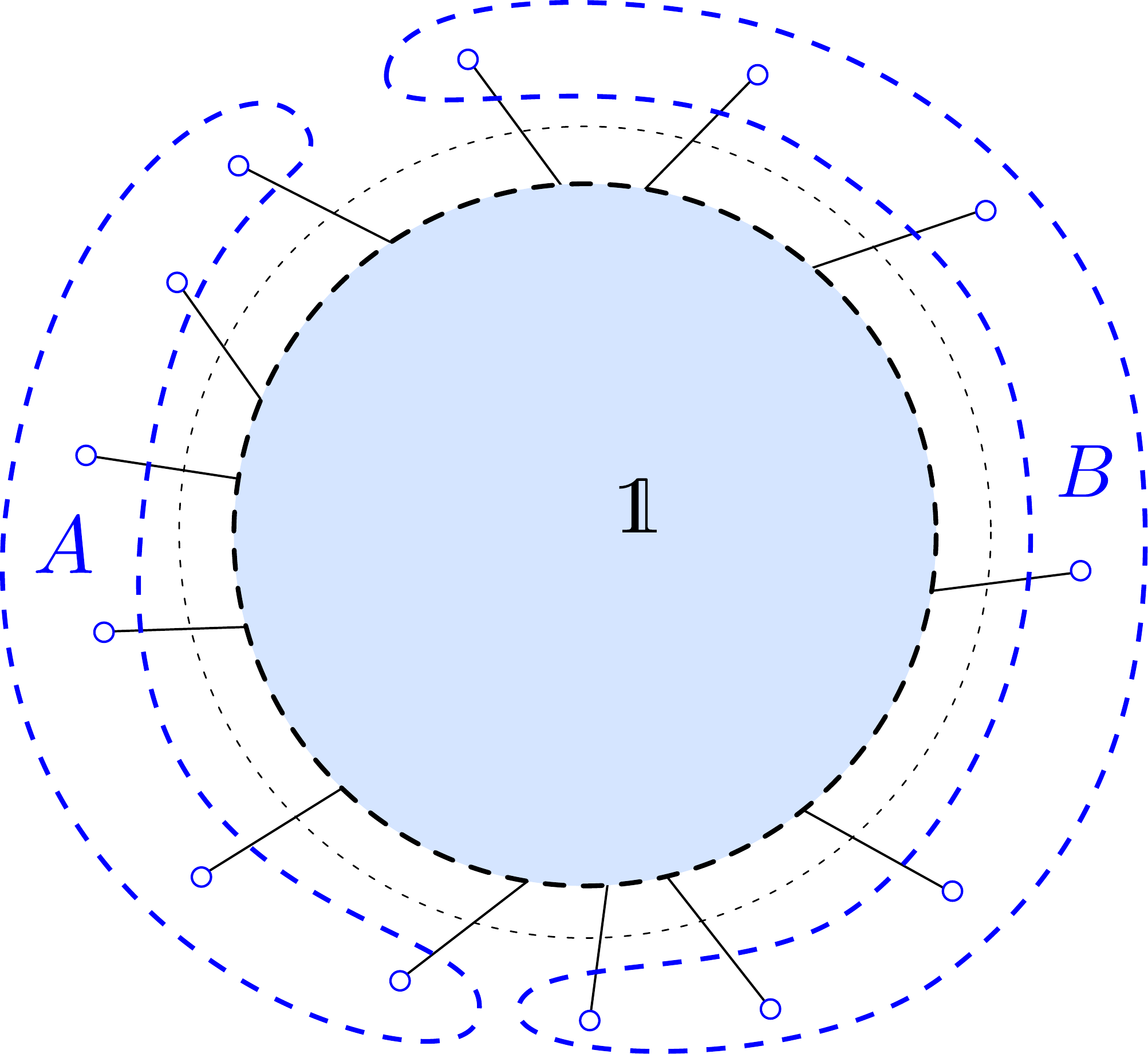}
\caption{The pattern without the minimal surface. Boundary condition: $A=\mathds{1}^{(A)}$ and $B=\mathds{1}^{(B)}$.}
\end{figure}

Finally, collecting the above results, we find that the leading contribution to $S_2$ is given as
\begin{eqnarray}
\mathbb{E}_T(\ex^{-S_2})\sim\frac{\mathbb{E}_T(Z_A^{(2)})}{\mathbb{E}_T(Z_0^{(2)})} = D^{-L_{\text{min}}}\left(1+O(D^{-1})\right) \qquad , 
\end{eqnarray}
which gives, when $D\gg 1$, 
\begin{eqnarray}
S_2 = L_{\text{min}}\ln D+O(D^{-1}) \qquad .
\end{eqnarray}
This is the Ryu-Takayanaki formula for $S_2$.

The $N$th R\'{e}nyi entropy $S_N$ can be calculated in a similar way, and with analogous result.

\section{Conclusions}
We have computed the Renyi entropy and derived the Ryu-Takanyagi entropy formula, for random tensor networks with an additional gauge invariance property. This is an interesting extension of existing derivations. On the one hand it shows the generality of the holographic behaviour of (random) tensor network states, thus their role in the entenglement/geometry correspondence, and confirming their interest also for applications in a quantum gravity context. On the other hand, the type of gauge symmetry we imposed is suggested by the correspondence with GFT and LQG states, and it is indeed required for the exact matching with the spin network states used in these quantum gravity approaches. Thus our results will facilitate the application of techniques from quantum gravity to quantum many-body systems (beyond the AdS/CFT framework), and the exploration of the same entanglement/geometry correspondence in new quantum gravity contexts.

\subsection*{Acknowledgements}
M. Zhang acknowledges support from the A. von Humboldt Foundation. 

%%%%%%%%%%%%%%%%%%%%%%%%%%%%%%%%%%%%%%%%

\appendix

\section{Structure of the matrix of $\mathbb{E}_T (\rho^{\otimes n})$}\label{Imatrix}
In this appendix we analyse the structure of the matrix in \Ref{eq:ETS} 
\begin{equation}
\mathbb{M}\equiv\sum_{\pi\in\mathcal{S}_n}\prod_{s=1}^n\delta_{\mu_{i(s)}\overline{\mu}_{\overline{i}(\pi(s))}}=\sum_{\pi\in\mathcal{S}_n}\mathds{P}_{\{i(s)\}\{\overline{i}(s)\}}(\pi;n,D^{v-1})
\end{equation}
The sum over $\mathds{P}(\pi;n,D^{v-1})$ is proportional to the projector operator $\mathds{P}_{\text{sym}}^{n,D^{v-1}}$ which projects vectors in $\mathbb{H}^{\otimes n}_{D^{v-1}}$ into its symmetric subspace. 
\begin{equation}
\mathds{P}_{\text{sym}}^{n,D^{v-1}}=\frac{1}{n!}\sum_{\pi\in\mathcal{S}_n}\mathds{P}(\pi;n,D^{v-1})
\end{equation}
Given a set of $\{i(s)\}=\{i(1),i(2),\cdots,i(n)\}$ and $\{\overline{i}(s)\}=\{\overline{i}(1),\overline{i}(2),\cdots,\overline{i}(n)\}$ where $i(s)$ and $\overline{i}(s)$ are from $0$ to $D-1$, there is a projector $\mathds{P}_{\text{sym}}^{n,D^{v-1}}$, which is a $D^{(v-1)n}\!\times\! D^{(v-1)n}$ matrix. Write $\Ref{eq:ETS}$ as a matrix, with $\{i(s)\}$ labeling its rows and $\{\overline{i}(s)\}$ labels its columns:
\begin{equation}
\mathbb{M}=n! 
\begin{pmatrix}
\mathds{P}_{\text{sym}}^{n,D^{v-1}} & \cdots & \mathds{P}_{\text{sym}}^{n,D^{v-1}}\\
\vdots & \ddots &\vdots\\
\mathds{P}_{\text{sym}}^{n,D^{v-1}} & \cdots & \mathds{P}_{\text{sym}}^{n,D^{v-1}}
\end{pmatrix}
\end{equation}
This matrix is a $D^n\!\!\times\!\! D^n$ block matrix.

The trace of $\mathbb{M}$ is
\begin{eqnarray}\label{eq:Mtr}
\tr\mathbb{M} &=& \sum_{\substack{\{i(s)\}\\ \{\overline{i}(s)\}}}\sum_{\pi\in\mathcal{S}_n}\prod_{s=1}^n\delta_{\mu_{i(s)}\overline{\mu}_{\overline{i}(\pi(s))}}\delta_{\mu_{i(s)}\overline{\mu}_{\overline{i}(s)}}\delta_{i(s)\overline{i}(s)} \nonumber\\
&=& \sum_{\substack{\{i(s)\}}}\sum_{\pi\in\mathcal{S}_n}\prod_{s=1}^n\delta_{\mu_{i(s)}\overline{\mu}_{i(\pi(s))}}\delta_{\mu_{i(s)}\overline{\mu}_{i(s)}}\nonumber\\
&=& \sum_{\substack{\{i(s)\}}}\sum_{\pi\in\mathcal{S}_n}\prod_{s=1}^n\delta_{\mu_{i(s)}\mu_{i(\pi(s))}}\nonumber\\
&=& \sum_{\substack{\{i(s)\}}} n! ~\tr\mathds{P}_{\text{sym}}^{n,D^{v-1}}\nonumber\\
&=& D^n D^{v-1}[n]
\end{eqnarray}
where we use the trace of the projector $\mathds{P}_{\text{sym}}^{n}$
\begin{equation}
\tr\mathds{P}_{\text{sym}}^{n,D^{v-1}}={D^{v-1}+n-1 \choose n}
\end{equation}

The matrix $\mathbb{M}$ can be written as a sum of matrices
\begin{eqnarray}
\mathbb{M} &=& \sum_{\pi\in\mathcal{S}_n}\prod_{s=1}^n\delta_{\mu_{i(s)}\overline{\mu}_{\overline{i}(\pi(s))}} \\ \nonumber
&=& \sum_{\{m(s)\}}\sum_{\pi\in\mathcal{S}_n}\prod_{s=1}^n\delta_{\mu_{[i(s)-m(s)]_D}\overline{\mu}_{\overline{i}(\pi(s))}}\delta_{i(s)\overline{i}(\pi(s))}\\
&=& \sum_{\{m(s)\}}\sum_{\pi\in\mathcal{S}_n}\prod_{s=1}^n\delta_{[\mu^{(s)}-m(s)D^{v-1}]_{D^v},\overline{\mu}^{(\pi(s))}}\\
&=& \sum_{\{m(s)\}}\sum_{\pi\in\mathcal{S}_n}\prod_{s=1}^n\prod_{a=1}^v\delta_{[\lambda_a(s)-m(s)]_D,\overline{\lambda}_a(\pi(s))}
\end{eqnarray}

Let us define a class of new matrices $\mathds{P}(\pi)_{\{m(s)\}}$ as
\begin{eqnarray}
\mathds{P}(\pi)_{\{m(s)\}} &\equiv& \prod_{s=1}^n\prod_{a=1}^v\delta_{[\lambda_a(s)-m(s)]_D,\overline{\lambda}_a(\pi(s))}\\ \nonumber 
&=& \prod_{s=1}^n\delta_{[\mu^{(s)}-m(s)D^{v-1}]_{D^v},\overline{\mu}^{(\pi(s))}}
\end{eqnarray}
Then matrix $\mathbb{M}$ becomes 
\begin{equation}\label{eq:MS}
\mathbb{M}=\sum_{\{m(s)\}}\sum_{\pi\in\mathcal{S}_n}\mathds{P}(\pi)_{\{m(s)\}} \qquad .
\end{equation}
When $m(s)=0$ for all $s$, $\mathds{P}(\pi)_{\{0\}}=\mathds{P}(\pi;n,D^v)$, the representation matrix of $\pi\in\mathcal{S}_n$ on $\mathbb{H}_{D^v}^{\otimes n}$.

%%%%%%%%%%%%%%%%%%%%%%%%%%%%%%%%%%%%%%%%

%\bibliographystyle{apsrev4-1}
%\bibliography{biblio}


\end{document}